\documentclass[twocolumn,showpacs,prl,floatfix]{revtex4}
\usepackage{graphicx}
\begin{document}
\title{Limits on diffuse fluxes of high energy extraterrestrial neutrinos
with the AMANDA-B10 detector}

\newcounter{foots}
\newcounter{notes}

\author{
J.~Ahrens$^{12}$,
X.~Bai$^{1}$, 
S.~W.~Barwick$^{6}$, 
R.~C.~Bay$^{5}$,
T.~Becka$^{12}$,  
K.-H.~Becker$^{13}$, 
E.~Bernardini$^{2}$,
D.~Bertrand$^{10}$,
A.~Biron$^{2}$, 
S.~Boeser$^{2}$
% ...
\vspace*{0.2cm}
}

\affiliation{$^1$Bartol Research Institute, University of Delaware, Newark, DE 19716}
\affiliation{$^2$DESY-Zeuthen,D-15735, Zeuthen, Germany}
\affiliation{$^3$Dept. of Technology, University of Kalmar, S-39182, Kalmar, Sweden}
\affiliation{$^4$Department of Physics, Stockholm University, SE-106 91 Stockholm, Sweden}
\affiliation{$^5$Dept. of Physics, University of California, Berkeley, CA 94720}
\affiliation{$^6$Dept. of Physics and Astronomy, University of California, Irvine, CA 92697}
\affiliation{$^7$Dept. of Physics, Pennsylvania State University, University Park, PA 16802}
\affiliation{$^8$Dept. of Physics, University of Wisconsin, Madison,  WI 53706}
\affiliation{$^9$Lawrence Berkeley National Laboratory, Berkeley, CA 94720}
\affiliation{$^{10}$Universite Libre de Bruxelles, Science Faculty CP230, Boulevard du Triomphe, B-1050, Brussels, Belgium}
\affiliation{$^{11}$Division of High Energy Physics, Uppsala University, S-75121, Uppsala, Sweden}
\affiliation{$^{12}$Institute of Physics, University of Mainz, Staudinger Weg 7, D-55099, Mainz, Germany}
\affiliation{$^{13}$Fachbereich 8 Physik, BUGH Wuppertal, D-42097 Wuppertal Germany}
\affiliation{$^{14}$Physics Dept., University of Wisconsin, River Falls,  WI 54022}
\affiliation{$^{15}$Vrije Universiteit Brussel, Dienst ELEM, B-1050, Brussels, Belgium}
\affiliation{$^{16}$Universit\'e de Mons-Hainaut, 19 Avenue Maistriau 7000,Mons, Belgium}
\affiliation{$^{17}$Dept. F\'{\i}sica, University Sim\'on Bol\'{\i}var, Caracas, Venezuela}
\affiliation{$^{18}$Blackett Laboratory, Imperial College, London SW7 2BW, United Kingdom}


\begin{abstract}
Data from the AMANDA-B10 detector taken during the austral winter of 1997
 have been searched for a diffuse flux of
high energy extraterrestrial muon-neutrinos, as predicted from, e.g., the
sum of all active galaxies in the universe. This search 
yielded no excess events above those expected from the background
atmospheric neutrinos, leading to upper limits on the extraterrestrial 
neutrino flux. For an assumed $\mathrm{E}^{-2}$ spectrum,
 a 90\% classical confidence level upper 
limit has been placed at a level 
 $\mathrm{E}^{2} \Phi(\mathrm{E}) = 8.4 \times 10^{-7}$
                      ${\mathrm{cm}^{-2}}\,{\mathrm{s}^{-1}}\,
{\mathrm{sr}^{-1}}   {\mathrm{GeV}}$,
(for a predominant neutrino energy range 6-1000 TeV)
 which is the most restrictive bound placed by any
neutrino detector. When specific predicted spectral forms are considered,
 it is found that some
are excluded. 

\end{abstract}

\maketitle

High energy extraterrestrial neutrinos 
 are believed to be produced in energetic accelerated environments through proton-proton or
proton-photon interactions via pion production and decay. Such an accelerator might be
the core of an active galaxy, powered by a supermassive black hole. In their
 pioneering work,
Stecker, Done, Salamon and Sommers \cite{SDSS91} calculated the
 expected diffuse flux of neutrinos from the
sum of all active galaxies and found that such a flux could be observable deep underground 
in a large 
 neutrino detector. Further predictions have followed (for a summary see for example
 the review of Learned and Mannheim \cite{ml}),
 and with  the construction and operation of the first high energy neutrino
detectors, the sensitivity has been reached to enable such predictions to be tested. Searches have
been made and limits
have been reported by the DUMAND \cite{DUMAND-SPS}, Frejus \cite{Frejus},
 Baikal$(\nu_e)$ \cite{baikal,baikal01}, MACRO \cite{MACRO2002} and AMANDA$(\nu_e)$ \cite{B10nue} neutrino 
detectors.  In this letter, we describe the search for 
high energy extraterrestrial neutrino-induced muons,
 using data collected during the austral winter of 1997
 with the AMANDA-B10 detector \cite{nature,amaatmos}, located in the antarctic 
ice cap at the South Pole station.

 The AMANDA (Antarctic Muon And Neutrino Detector Array)
 telescope detects high-energy muon-neutrinos by observing Cherenkov light from muons
resulting from neutrino interactions in the ice surrounding, or the  rock, below
the detector.  
While extraterrestrial neutrinos will produce high-energy muons from all arrival
directions, those coming from \emph{above} 
the detector will be very difficult to separate from
the overwhelming flux of downward-going cosmic-ray induced atmospheric muons. The majority of
these muons are rejected by only accepting upward going neutrino-induced
 muons; the earth filters out muons
produced in the atmosphere on the other side of the planet. There is a small remaining
flux of mis-reconstructed events which is removed by quality cuts that leave only
well reconstructed events. After the atmospheric muons are removed, there will 
remain a flux of upward going muons from
 cosmic ray induced atmospheric neutrinos that have penetrated the earth and
interacted near the detector. This relatively well 
understood  neutrino flux is a background to the search but has been
 used to verify
the performance of the detector \cite{nature,amaatmos}.
The separation of the  extraterrestrial
neutrino-induced muons from the atmospheric neutrino-induced muons
 is based on the expected energy spectrum of the detected
muons.
  Typically, 
a model of an extraterrestrial source
 of neutrinos has a harder spectrum (e.g. $\sim \mathrm{E^{-2}}$) \cite{ss95}
than that of the atmospheric neutrinos $(\sim\mathrm{E^{-3.7}})$ \cite{Lipari93,gh}. 
After accounting for neutrino interaction and muon propagation, this energy
difference carries over to produce a harder muon energy spectrum for the
extraterrestrial neutrino-induced muons near the
detector. The energy of the muon is not measured directly, but  
 more energetic muons tend to produce more Cherenkov light and thus 
more hit optical modules  in the detector, and this observable, the
 channel multiplicity ($\mathrm{N_{ch}}$), is
used as the primary separator of higher energy extraterrestrial neutrino-induced muon
 events from the background of lower energy atmospheric 
neutrino-induced muons. 

\begin{figure}[h]
\begin{center}
\resizebox{3.5 in}{!}
{\includegraphics{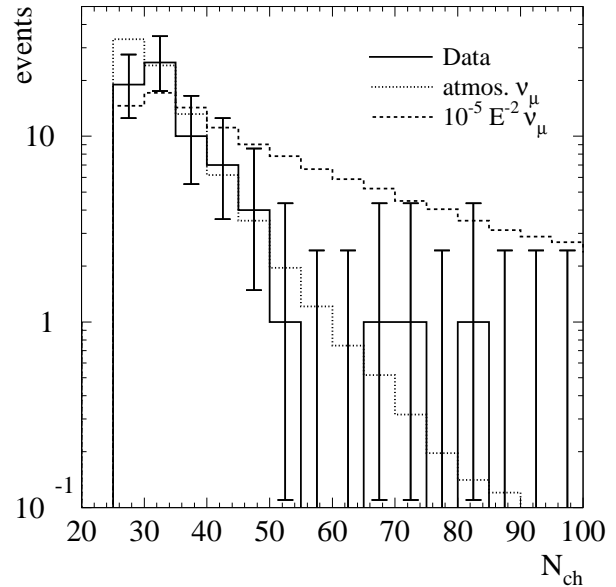}}
\caption{\label{fig:diffnch} {Channel multiplicity distribution after final cuts, showing
the expected excess of events from an $\mathrm{E^{-2}}$ spectrum at the higher multiplicities.}}
\end{center}
\end{figure}

In this analysis, the event selection cuts were designed to retain high energy
track-like events \cite{leu2001}. The detector simulation has changed from that used in the atmospheric
neutrino analysis \cite{amaatmos}. A new muon propagation code \cite{mmc} was used, which
 accounts for all relevant stochastic light emission from the
muons. The depth-dependent optical properties of the fiducial ice were
determined using atmospheric muons as a calibration source.

Before the energy sensitive channel multiplicity cut was finally  applied, 69
 events remained in the data sample,
whereas a full simulation of the detector response to the  atmospheric neutrino  
(Lipari \cite{Lipari93}) flux (neglecting neutrino oscillations, which would reduce the prediction
by only a few percent)
predicts 85 events for the 130 days of live time.
 The absolute difference in the numbers of events is consistent with 
Poisson fluctuations, or 
with the $\pm 25\%$ \cite{gh} uncertainty in the atmospheric neutrino flux, or with uncertainties in 
the simulation efficiencies.   The distribution of the data and atmospheric simulation
are shown in  figure \ref{fig:diffnch}. The error bars on the data are 90\% unified confidence intervals \cite{fc98}
for the fixed but unknown value of the mean rate $\mu=\mu_b+\mu_s$ for each bin. Only one bin
($\mathrm{N_{ch}=25-30}$) has a background prediction inconsistent with the confidence interval. 
 More specificially, a 
generalised likelihood ratio test of the shape of the atmospheric neutrino hypothesis
 as the parent
distribution of the data yields a chance probability of 20\%, which is too large to reject the shape
of the  atmospheric
neutrino hypothesis.
We choose to treat the rate of
 observed atmospheric neutrinos as a constraint on the overall detector efficiency and then carry 
through an efficiency uncertainty from the atmospheric neutrino flux prediction and Poisson error on the observed
rate.  Therefore, to calibrate the overall detector sensitivity, we take the 69 events as
the best-fit estimate of the number of atmospheric neutrinos and rescale all efficiencies by a factor
 69/85. This is conservative, since 
if the first bin discrepancy was due e.g. to a simulation effect, then no
 renormalisation would be needed, and the limits would improve slightly.
 We combine  the Poisson error on the observed rate of atmospheric neutrinos 
with the theoretical flux uncertainty (taken as a uniform probability distribution
 centred about the best-fit flux $\hat\Phi$ and extending to $\pm 0.25\hat\Phi$) to compute the correlations
between the background and efficiency for later use in the probability distribution function used in 
the confidence interval construction.  To incorporate these systematic
uncertainties in the  efficiencies
into the limit calculations, 
 we follow the prescription
of Cousins and Highland \cite{ch}, as implemented by Conrad {\em et al.} \cite{BCHD} 
with the unified Feldman-Cousins
 ordering and
improved by a more appropriate choice
 of the likelihood ratio test \cite{gch-sys}. We also  report all limits and sensitivities
 with and without the
assumed uncertainty. 

 In addition to the data and atmospheric neutrino prediction,
 figure \ref{fig:diffnch} also shows
  the prediction for an $\mathrm{E^{-2}}$ signal flux at a
level 
 $\mathrm{E}^{2}\, \Phi(\mathrm{E}) = 10^{-5}\, {\mathrm{cm}^{-2}}\,{\mathrm{s}^{-1}}\,
{\mathrm{sr}^{-1}} \,  {\mathrm{GeV}}$, a flux that would have  been readily detected.
Setting a limit  on a flux $\Phi(\mathrm{E})$ involves determining an experimental
signal event upper
limit $\mu(\mathrm{n_{obs},n_b})$, which is a function of the number of observed events,
$\mathrm{n_{obs}}$, and expected background, $\mathrm{n_b}$, after the cuts are applied. A
simulation chain accounting for neutrino absorption,
interaction and neutral current regeneration, muon propagation and detector response gives
 the number of signal events, $\mathrm{n_s}$, expected from the
source flux  $\Phi(\mathrm{E})$. The limit on the source flux will then be
$\Phi_{\mathrm{limit}}\mathrm{(E)} = \Phi(\mathrm{E}) \times {\mu(\mathrm{n_{obs},n_b})}/{\mathrm{n_s}}$.
The choice of final cut is optimised before examining the data by 
 minimising the average
 ``model rejection factor'' (MRF)
 ${\bar{\mu}(\mathrm{n_{obs},n_b})}/{\mathrm{n_s}}$ \cite{gchkr2000}, where the
as yet unknown experimental
event limit $\mu(\mathrm{n_{obs},n_b})$ is replaced by the  
 {\em average}
 upper limit ${\bar{\mu}(\mathrm{n_b}})$ \cite{fc98}. Over an
ensemble of hypothetical repetitions of the experiment, this choice of
cut will lead to the best average limit $\bar{\Phi}_{\mathrm{limit}}\mathrm{(E)}$.

\begin{figure}[htb]
\begin{center}
\resizebox{3.5 in}{!}
{\includegraphics{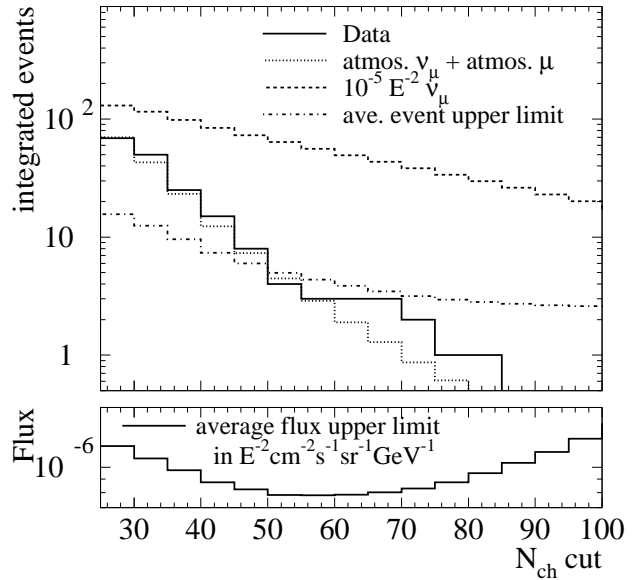}}
\caption{\label{fig:mrpintegral} {Integrated distributions of event numbers as a
function of the multiplicity cut (top plot). The minimum in the  flux average upper limit (bottom)
is found by minimising the ratio of the average event upper limit to the 
 expected $\mathrm{E^{-2}}$ signal.}}
\end{center}
\end{figure}
 
\begin{figure}[htb]
\begin{center}
\resizebox{3.5 in}{!}
{\includegraphics{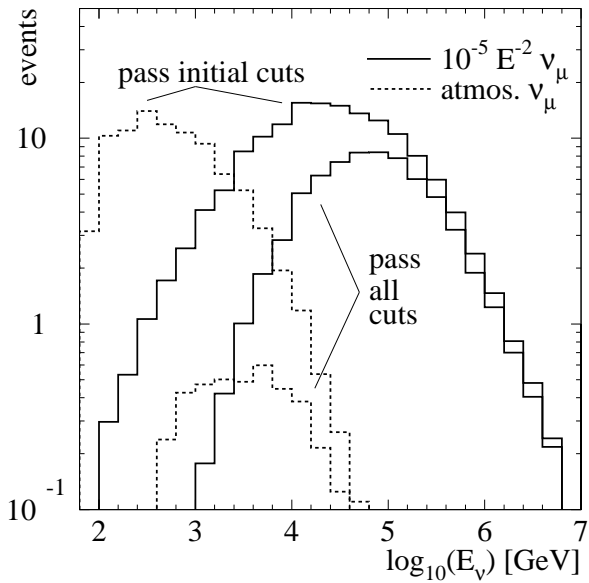}}
\caption{\label{fig:nuenfinal} {Energy spectrum of the incident
atmospheric (dashed lines) and $\mathrm{E^{-2}}$ (solid lines) neutrinos for
events that pass the initial cuts (thin lines), and have channel multiplicity greater than the
optimum cut of 54 channels (thick lines).}}
\end{center}
\end{figure}

 The integrated channel multiplicity
distribution is shown in
 figure \ref{fig:mrpintegral}.
Also shown is the
90\% confidence level
Feldman-Cousins average upper limit which is a
 function of the expected background. The optimal
cut is the one where the model rejection factor  $ {\mu(\mathrm{n_b}})/{\mathrm{n_s}} $ is 
minimised. Figure 
\ref{fig:mrpintegral} also shows the average flux upper limit ($\mathrm{E^2}{\Phi} \times \mathrm{MRF}$)
 as a function of the 
choice of multiplicity cut. The minimum flux limit occurs at a cut of
 $\mathrm{N_{ch}} \ge 54$, where 
we expect $\mathrm{n_b}=3.06$ and an average signal event upper limit of 4.43 ignoring the uncertainties in the
efficiency and background, and
4.93 when the uncertainties are included.
 The $10^{-5} \mathrm{E^{-2}}$ 
signal flux would produce 56.7 events.    This leads
to corresponding expected average limits  on the source flux of 
 $\mathrm{E}^{2} {\bar{\Phi}}_{\mathrm {90\%}}(\mathrm{E})  = 7.8 \times  10^{-7}  \,{\mathrm{cm}^{-2}}\,{\mathrm{s}^{-1}}\,
{\mathrm{sr}^{-1}}  \, {\mathrm{GeV}}$ (excluding uncertainties), and $8.7 \times  10^{-7}  \,{\mathrm{cm}^{-2}}\,{\mathrm{s}^{-1}}\,
{\mathrm{sr}^{-1}}  \, {\mathrm{GeV}}$ (including uncertainties).

We note that the expected overall flux limit is
relatively insensitive to the choice of cut, with a broad minimum seen in figure 
\ref{fig:mrpintegral} in the range
of multiplicities 50-70. 
We now apply this optimal multiplicity cut to the data, and find that 3
 events remain. Ignoring the systematic uncertainties gives an event limit of 4.36 and a
flux upper limit of 
$\mathrm{E^{2}} \Phi_{\mathrm{90\%}}(\mathrm{E}) = 7.7 \times  10^{-7}  \,{\mathrm{cm}^{-2}}\,{\mathrm{s}^{-1}}\,
{\mathrm {sr^{-1}}}   {\mathrm{GeV}}$. Including the systematic uncertainties leads to an event limit
of 4.75 and  our
final flux limit on an $\mathrm{E^{-2}}$ spectrum of
$\mathrm{E^{2}} \Phi_{\mathrm{90\%}}(\mathrm{E}) = 8.4 \times  10^{-7}  \,{\mathrm{cm}^{-2}}\,{\mathrm{s}^{-1}}\,
{\mathrm {sr^{-1}}}   {\mathrm{GeV}}$.

Figure \ref{fig:nuenfinal} shows the neutrino energy spectrum
 of the simulated events before and after the 
multiplicity cut of 54 channels, for both atmospheric neutrinos and neutrinos from
an $\mathrm{E^{-2}}$ spectrum. The multiplicity cut corresponds to a sensitive
 energy range of 6-1000 TeV, which contains 90\% of the expected $\mathrm{E^{-2}}$ signal. The 
peak response energy is just below 100 TeV.

\begin{figure}[htp]
\begin{center}
\resizebox{3.5 in}{!}
{\includegraphics{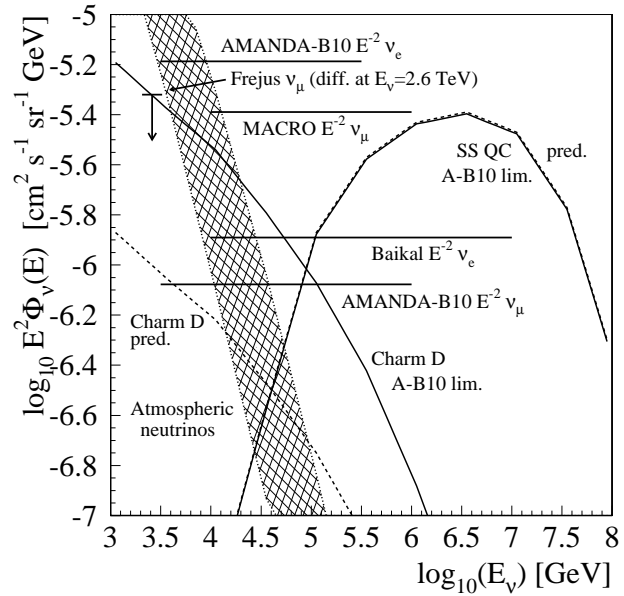}}
\caption{\label{diffuse_mrf_plot_summary}Summary of experimental 90\% classical confidence level
 flux limits from various
detectors assuming an $\mathrm{E^{-2}}$ spectrum. From top: AMANDA-B10 ($\nu_e$) \cite{B10nue},
 Frejus \cite{Frejus}, MACRO \cite{MACRO2002},
 Baikal \cite{baikal01} and
AMANDA-B10 ($\nu_{\mu}$) (this work). The background atmospheric neutrinos \cite{Lipari93} are
indicated by the hashed region representing the angular dependence of the flux.
 Also shown are the predicted fluxes (dashed), and 
AMANDA-B10 experimental flux limits (solid) for a diffuse AGN prediction (SSQC \cite{ss95} -- 
nearly overlapping dotted and dashed curves -- MRF = 0.98) 
and for one prediction of prompt charm neutrino production in the earth's atmosphere \cite{ZHV92}.
Since most events will originate from neutrinos near the peak of the detector sensitivity
 ($\mathrm{E_{\nu}\sim 10^5\, GeV}$), the limits at
that point for different spectral shapes are similar.
}
\end{center}
\end{figure}

Just as a limit was placed on an assumed
 $\mathrm{E^{-2}}$ spectrum, limits can be placed on any neutrino flux prediction, and 
we consider a sample of predictions that are near the limit-setting 
capability of the detector \mbox{($\mathrm{MRF}\sim 1$)}.
For each case, we optimise the final $\mathrm{N_{ch}}$ cut to minimise the expected 
average flux
upper limit, then 
compare the expected number of extraterrestrial neutrino events after the cuts to the observed 
event limit; those predictions that produce expected event numbers greater 
than the observed event limit are 
excluded at the stated classical confidence level. The results of these calculations are shown 
in table \ref{fluxlimittable} and in 
 figure \ref{diffuse_mrf_plot_summary}. For each flux we again report two sensitivities and limits
 - one assuming no systematic uncertainties  and the second including systematic uncertainties.
  We find that the predictions of Szabo and Protheroe (SPH92L \cite{SPH92},
 P96p$\gamma$pp \cite{P96}) are excluded.
The quasar core (SSQC) prediction of Stecker and Salamon \cite{ss95} is just 
excluded ($\mathrm{MRF}=0.98$), but the blazar jet (SSBJ)
prediction is not. The limit of the original Stecker,
Done, Salamon and Sommers flux \cite{SDSS91} (SDSS) is a factor of 2 above the prediction and therefore the prediction is not excluded. 

Maximal mixing from neutrino oscillations \cite{sk,sno} 
between $\nu_{\mu}$ and $\nu_{\tau}$ during  propagation to the earth
 would lead to limits 
a factor of two greater than quoted here due to  the loss
of half the expected $\nu_{\mu}$ flux into the $\nu_{\tau}$ state. However,
 some of these $\nu_{\tau}$ would regenerate $\nu_{\mu}$ in the earth ($\nu_{\tau}\rightarrow\tau
\rightarrow\nu_{\mu}$)
 lessening the effect \cite{hs,bck}. 

We also place a limit on a model of prompt charm
induced neutrinos \cite{ZHV92} (ZHV92) in the earth's atmosphere
and find that the detector sensitivity is about a factor of 4 away from excluding the prediction.  More
 recent predictions are even further below the sensitivity of the detector \cite{costa}.

Since most events will originate from neutrinos near the peak of the detector sensitivity
 $\mathrm{E_{\nu}\sim 10^5\, GeV}$, the limits at
that point for the three  different spectral shapes ($\mathrm{E^{-2}}$, SSQC and Charm D) are similar,
 as seen in \hbox{figure \ref{diffuse_mrf_plot_summary}}.

\begin{table*}[htb]

\caption{\label{fluxlimittable}
Flux limits calculated for individual models of diffuse neutrino emission. The optimal $\mathrm{N_{ch}}$
cut, expected background and signal for each model are shown. The average upper limit ($\bar{\mu}(\mathrm{n_b})$)
and average model rejection factor (${\bar{\mu}(\mathrm{n_b})}/{\mathrm{n_s}}$) are shown
with and without the inclusion of systematic uncertainties. Finally the experimental limits (event limit
 ${\mu_{\mathrm{o}}}\equiv \mu(\mathrm{n_{o},n_b}$)) and model rejection factor (${{\mu_{\mathrm{o}}}}/{\mathrm{n_s}}$) are given
for both systematic uncertainty assumptions.}
\vspace{0.5cm}

\begin{tabular}{lcccccccccccc} \hline \hline

\multicolumn{4}{c}{} & \multicolumn{4}{l}{Sensitivities} &
\multicolumn{5}{l}{Experimental limits} \\

\multicolumn{4}{}{} & \multicolumn{2}{l}{No sys. uncer.} & 
\multicolumn{2}{l}{Sys. uncer. inc.} & & \multicolumn{2}{l}{No sys. uncer.} & 
\multicolumn{2}{l}{Sys. uncer. inc.}\\

Flux & nch cut & $\mathrm{n_b}$ & $\mathrm{n_s}$ 
    & $\bar{\mu}(\mathrm{n_b})$  &
 $\frac{\bar{\mu}(\mathrm{n_b})}{\mathrm{n_s}}$ & $\bar{\mu}(\mathrm{n_b})$
       & $\frac{\bar{\mu}(\mathrm{n_b})}{\mathrm{n_s}}$& 
$  \mathrm{n_{o}}$ & ${\mu_{\mathrm{o}}}$  & 
 $\frac{{\mu_{\mathrm{o}}}}{\mathrm{n_s}}$  & ${\mu_{\mathrm{o}}}$  & 
 $\frac{{\mu_{\mathrm{o}}}}{\mathrm{n_s}}$   \\   \hline

   $10^{-6}\mathrm{E^{-2}}$      & 54 & 3.06 & 5.67 & 4.43  &  0.781 &  4.93 &  0.869 & 3 & 4.36& 0.769 & 4.75& 0.838\\

\hline
   SDSS \cite{SDSS91}             & 73 & 0.69 & 2.42  & 3.01  &  1.240 &    3.38 &  1.397 & 2 & 5.22   & 2.157  & 5.61 & 2.318\\
\hline
    SPH92L \cite{SPH92}           & 58 & 2.12 & 12.66 & 3.97  & 0.314   &  4.33  & 0.342 & 3    &  5.30     & 0.419    &  5.69 & 0.449  \\
\hline
SSQC \cite{ss95}                & 71 & 0.80 & 5.59 & 3.11  & 0.556    &  3.45  & 0.617 & 2    &  5.11  & 0.914     &  5.50 & 0.984 \\
\hline
SSBJ \cite{ss95}                & 57 & 2.36 & 4.29 & 4.13 &  0.963   &  4.50  & 1.049 &3    & 5.06  & 1.179     &  5.45 & 1.270   \\
\hline
 P96p$\gamma$pp \cite{P96}      & 49 & 4.83 & 21.95 & 5.11  & 0.233   &  5.90  & 0.269 &4   & 3.76    & 0.171     &  4.54 & 0.207  \\
\hline
  ZHV Charm D \cite{ZHV92}        & 41 & 10.9 & 2.58 & 6.97  &   2.702 &  8.42 & 3.264 & 14 & 10.60& 4.109 & 12.31 & 4.771 \\  

\hline
\hline

\end{tabular}
\end{table*}

 The limits presented in this letter, based on the first real-time year
of operation of the AMANDA-B10 detector, are the  strongest placed to date on 
extraterrestrial diffuse neutrino fluxes. Since that year, we estimate that
about 10 times the exposure has been achieved in total with AMANDA-B10 (1997-99) and the
 expanded AMANDA-II detector (2000-present). We anticipate this combined data
set to have a limit-setting potential more than three times better than the results presented
here. 

We acknowledge the support of the following agencies: National
Science Foundation--Office of Polar Programs, National Science
Foundation--Physics Division, University of Wisconsin Alumni Research
Foundation, Department of Energy, and National Energy Research
Scientific Computing Center (supported by the Office of Energy
Research of the Department of Energy), UC-Irvine AENEAS Supercomputer
Facility, USA; Swedish Research Council, Swedish Polar Research
Secretariat, and Knut and Alice Wallenberg Foundation, Sweden; German
Ministry for Education and Research, Deutsche Forschungsgemeinschaft
(DFG), Germany; Fund for Scientific Research (FNRS-FWO), Flanders
Institute to encourage scientific and technological research in
industry (IWT), and Belgian Federal Office for Scientific, Technical
and Cultural affairs (OSTC), Belgium.  D.F.C. acknowledges the support
of the NSF CAREER program.

\end{document}